# Probing the Electronic Structure of Graphene Near and Far from the Fermi Level via Planar Tunneling Spectroscopy

John L. Davenport[1], Zhehao Ge[1], Junyan Liu[1], Carlos Nuñez-Lobato[1], Seongphill Moon[3,4], Zhengguang Lu[3,4], Eberth A. Quezada[1], Kaitlin Hellier[1], Patrick G. LaBarre[1], Takashi Taniguchi[2], Kenji Watanabe[2], Sue Carter[1], Arthur P. Ramirez[1], Dmitry Smirnov[3], Jairo Velasco Jr.[1*]

[1]Department of Physics, University of California, Santa Cruz, California 95064, USA
[2]National Institute for Materials Science, 1-1 Namiki, Tsukuba, 305-0044, Japan
[3]National High Magnetic Field Laboratory, Tallahassee, FL 32310, USA
[4]Department of Physics, Florida State University, Tallahassee, FL 32306, USA

*Email: jvelasc5@ucsc.edu

Scanning tunneling spectroscopy (STS) has yielded significant insight on the electronic structure of graphene and other two-dimensional (2D) materials. STS directly measures a fundamental and directly calculable quantity: the single particle density of states (SPDOS). Due to experimental setup limitations, however, STS has been unable to explore 2D materials in ultra-high magnetic fields where electron-electron interactions can drastically change the SPDOS. Recent developments in the assembly of heterostructures composed of graphene and hexagonal boron nitride have enabled a device-based alternative to potentially overcome these roadblocks. Thus far, however, these nascent efforts are incomplete in analyzing and understanding tunneling spectra and have yet to explore graphene at high magnetic fields. Here we report an experiment **at magnetic fields up to 18 T** that uses graphene tunneling field effect transistors (TFETs) to establish a clear benchmark for measurement and analysis of graphene planar tunneling spectroscopy. We acquire **gate tunable tunneling spectra of graphene** and then use these data and electrostatic arguments to develop a systematic analysis scheme. This analysis reveals that TFET devices directly probe electronic structure features near and far from the Fermi level. In particular, our study yields identification of the Dirac point and numerous Landau levels as they fill and empty with charge *via* application of a gate voltage. Our work demonstrates that TFET devices are a viable platform for directly probing the electronic structure of graphene and other 2D materials in high magnetic fields, where novel electronic states emerge.

Scanning tunneling spectroscopy (STS) is a powerful tool used to directly probe the electronic structure of graphene nanodevices.[1,2] Physical phenomena related to electronic correlations are prevalent in these nanodevices in ultra-high magnetic fields and lead to drastic change in the graphene single particle density of states (SPDOS). However, STS measurements require extreme mechanical stability and are thus incompatible with experimental setups that incorporate ultra-high magnetic fields (> 15 T).[3] While capacitance spectroscopy and electronic transport do not require the same mechanical stability as STS, they are limited to measure the SPDOS of a material at its Fermi Level.[4,5] The ability to examine electronic structure features at high magnetic fields both near and far from the Fermi level would enable a more comprehensive comparison with theoretical models including those that account for electron-electron interactions.[6] Previous works have realized device based alternatives for STS *via* heterostructures composed of graphene and hexagonal boron nitride (hBN). These tunneling field effect transistors (TFET) have been used to study phonons[7,8] and Andreev bound states.[9,10] Landau Levels have also been probed but indirectly and at low magnetic fields.[11] These recent works, however, have been inconsistent in treating the probe-sample coupling, a factor that can drastically alter quantitative conclusions drawn from tunneling spectra. A more complete account of the probe-sample coupling in graphene TFETs is necessary for understanding graphene planar tunneling spectroscopy.

In order to develop a tunneling spectra benchmark for graphene TFETS, we acquired gate tunable planar tunneling spectra and developed a new systematic method for analyzing/interpreting such spectra. Consequently, our study enables the first direct measurement of numerous graphene electronic structure features at B = 18 T as they fill and empty with charge. As schematized in Fig. 1a, the TFET device employed for our study consists of a van der Waals heterostructure with graphene (gray) sandwiched in between a bottom thick hBN flake (blue) and a top few-layer hBN flake (also blue). In this structure, the few-layer hBN flake acts as a tunneling barrier between the graphene sample and a top lying polycrystalline Cr/Au electrode (gold).[12] The thick hBN platform is used to screen graphene from the influence of extrinsic factors such as charge impurities.[13] The entire heterostructure rests on a 300 nm $SiO_2$ substrate. An optical micrograph of a completed TFET device is shown in Fig. 1b. Numerous probe electrodes that rest on top of a five-layer hBN flake are outlined in orange. One electrode contacts graphene directly and is outlined in black. The exposed (concealed) region of the graphene is denoted by a solid (dashed) line. When a sample bias $V_b$ (which is equivalent to the negative of a probe bias $V_{probe}$) is applied between the probe and graphene, charges tunnel through the barrier and into available states in graphene. The resulting tunneling conductance $dI/dV_b$ measured across the junction is proportional to the SPDOS of graphene at energy $eV_b$.[14] A gate voltage $V_G$ is applied to the p-doped silicon and is used to capacitively tune graphene's charge density $n$ and enables gate tunable $dI/dV_b$ measurements.[1]

We first obtain gate tunable $dI/dV_b$ spectra at 4.2 K from our TFET device, an example of which is plotted as a function of $V_b$ and $V_G$ in Fig. 1c. The data show a prominent, horizontal ~0.130 V dark stripe centered at $V_b = 0$ V. Additional features are present in the data but are obscured by the presence of this horizontal stripe. To access the other features from Fig. 1c we take the numerical derivative of these data and plot this in Fig. 1d as $dG/dV_G$ ($V_G$, $V_b$). Two additional features that exhibit a $dG/dV_G$ intensity close to zero become immediately apparent: (i) A narrow diagonal white line that moves from the bottom left to the top right of the $dG/dV_G$ ($V_G$, $V_b$) map, crossing the horizontal stripe feature from Fig. 1c (now white) at $V_G = 20$ V and reappearing at $V_G = 40$ V; and (ii) Another narrow white line that behaves sublinearly with respect to $V_G$ and crosses the horizontal white band at the same points in $V_G$ as (i) but changes concavity upon crossing the horizontal white stripe. Features (i) and (ii) are denoted by a black dotted line and dashed line, respectively.

To gain further insight on the two $V_G$-dependent features in Fig. 1c we extract several profiles within Fig. 1c along diagonal lines that are parallel to feature (i). Each of these line profiles is displayed in Fig. 1e as a $dI/dV_b(V_b)$ spectrum with a corresponding color from Fig. 1c and with a vertical offset. The spectra show a strong 0.130 V-wide suppression in tunneling $dI/dV_b$ centered around $V_b = 0$ V that is $V_G$-independent. This feature corresponds to the dark horizontal stripe in Fig. 1c and has been seen previously in tunneling spectroscopy and STS studies of graphene/SiO$_2$ devices[1, 15, 16] and graphene/hBN heterostructures.[7, 17] Its nature is well established and is attributed to phonon assisted inelastic tunneling into graphene.[7, 17] An additional feature is visible in each of the line profiles presented in Fig. 1e as a minimum adjacent to the prominent inelastic tunneling feature mentioned above. The position of the minima in Fig. 1e depends on the location of the respective diagonal lines in Fig. 1c. For example, the red line profile displays the minimum at positive $V_b$, whereas the purple line profile displays the minimum at negative $V_b$. For line profiles that are in between (the red and purple profiles) the minimum shifts monotonically from positive to negative $V_b$.

Previous STS and TFET studies have shown features akin to either feature (i), feature (ii),[2,3] or both.[4,5] Notably, the work by Zhao, et. al. attributes the two separate features to tunneling channels available at the chemical potential of the probe $\mu_p$, and the chemical potential of graphene $\mu_g$, which is modulated by $V_b$. Importantly, this interpretation implies that spectra taken at constant $V_G$ do not correspond to a constant $n$.[6] In our experiment a similar systematic consideration of the graphene $n$ with respect to $V_b$ modulation is crucial for understanding the observed spectral features and relating them to prior STS work. This is because the TFET device geometry has a strong capacitive coupling between the top lying tunneling probe and graphene due to their proximity. In effect, graphene's $n$ (and chemical potential $\mu_g$) shift if $V_G$ or $V_b$ is modulated. This effect has two significant consequences: (1) There is an additional contribution to the tunneling $dI/dV_b$ that is related to tunneling at $\mu_g$; and (2) Applying a $V_b$ shifts spectral features away from their expected energies, thus complicating the quantitative interpretation of these spectral features. The origin of effect (1) was predicted by Malec, et. al., and addressed in the STS work by Zhao, et. al. A discussion on this effect in our TFET structure is provided in the supplementary information. Effect (2) has been identified in the literature as band shifting,[18] but is often neglected in graphene STS works,[17,19] including recent works on magic angle twisted bilayer graphene.[20,21]

In order to extract accurate quantitative electronic structure information from our TFET spectra, it is necessary to systematically account for effect (2). Motivated by the similarities between diagonal line profiles of Fig. 1e to gate tunable STS of graphene/hBN heterostructures,[2] we shear $dI/dV_b(V_b, V_G)$ maps, thus transforming diagonal line profiles onto vertical lines in the resulting $dI/dV_b(V_b, n)$ maps. A map resulting from this shearing process is shown in Fig. 2a. It reveals feature (i), discussed above, as a vertical dark band (indicated with a yellow arrow). Additionally, feature (ii) is also visible, it is denoted as a white dashed line in the same $dI/dV_b(V_b, n)$ map. This transformation enables the study of planar tunneling spectroscopy at well-defined $n$. More details on the technical aspects of the shear transformation and determination of $n$ can be found in the supplementary information.

Figure 2b shows a $dI/dV_b(V_b, n)$ map of the same device in a perpendicular magnetic field $B = 8\ T$. These data reveal features that are present in Fig. 1c and new features as well. The horizontal dark stripe seen in Fig. 1c (and Fig. 2a) is observed again in Fig. 2b. In addition, new peaks appear that follow the path carved by feature (ii) in Fig. 2a. Interestingly, the trajectory of these new peaks changes significantly as $n$ approaches 0, exhibiting a characteristic step-like behavior with $n$. The onset of this step-like behavior, denoted by a black arrow, is apparent on the bottom half of the map, where $V_b < 0$ V. Figure 2c shows a comparison of constant $n$ tunneling spectra when $B = 0\,T$ (black) and $B = 8\,T$ (blue) at $n = -1.8 \times 10^{12}$ cm$^{-2}$. Both spectra show the inelastic tunneling feature around $V_b = 0$ V, and the blue spectrum shows a broad peak in the location of the Dirac point in its $B = 0\,T$ counterpart. Several additional peaks are also present in the blue $dI/dV_b(V_b)$ spectrum.

To further study the spectral features that emerge at $B = 8\,T$, we measured TFET devices at high magnetic fields. Figures 3a-c show sheared $dI/dV_b(V_b, n)$ maps of a graphene TFET for $B = 8\,T$, $B = 13\,T$ and $B = 18\,T$, respectively. The vertical dashed line cuts in each of these panels is taken along the charge neutrality point and corresponds to a $dI/dV_b(V_b)$ spectrum of matching color in Fig. 3d. with $B = 8\,T$ (orange), $13\,T$ (blue), and $18\,T$ (black). All of the three spectra show the inelastic tunneling feature discussed above and adjacent sharp peaks in $dI/dV_b$ intensity. Additionally, the spectra show a peak indicated by a pink arrow that shifts to the left as $B$ is increased. This peak is visible as a stair-like dispersing band in Fig. 3a, and it appears again in Figs 3b and 3c at lower values of $V_b$. Notably, the length $\Delta n$ of this step-like feature increases with larger magnetic fields, eventually forming a prominent staircase pattern for $B = 18\,T$ (Fig. 3c).

We now use our data that has undergone the shear transformation to accurately extract electronic structure parameters such as the Fermi velocity $v_F$ and the state degeneracy factor $g$. We first look at the $v_F$ for $B = 0\,T$ by determining the position (in $V_b$) of feature (ii) from $dI/dV_b(V_b, n)$ maps shown in Fig. 2a. The energies of feature (ii), which should correspond to the energy of the Dirac point ($E_{DP}$) with respect to $\mu_g$, are plotted as a function of fixed $n$ in Fig. 4a with the phonon assisted inelastic tunneling energy $E_{ph}$ subtracted.[1] Because of graphene's linear density of states and 2D nature, at low energies $E_{DP} = \hbar v_F \sqrt{n\pi}$ .[25] This expectation is plotted in Fig. 4a as a black solid line. The data and theoretical dispersion match, thus indicating that our TFET spectra are directly measuring $E_{DP}$, including when it is far away (> 0.2 eV) from the Fermi Level.[1,17] To further support this claim we include in the Supporting Information a simulation of $dI/dV_b$ that accounts for the effect of graphene's quantum capacitance in determining $n$.

Having quantitively established the validity of the shear transformation, we now apply a similar analysis to $dI/dV_b(V_b, n)$ maps for $B \neq 0\ T$ (Fig. 2b). This analysis enables an additional verification of the graphene linear dispersion and an examination of its behavior at different $n$. In the presence of a large $B$, charges in a 2D electron gas undergo cyclotron orbits that coalesce to form Landau levels (LL). Characteristic of materials with a Dirac dispersion, the energies of these LLs are given by $E_{LLN} - E_{LL0} = sgn(N)v_F\sqrt{2e\hbar|N|B}$ , where $E_{LLN}$ is the energy of the $N^{th}$ LL and $E_{LL0}$ is the energy of the zeroth LL. We use constant $n$ line profiles taken from Figs. 2a and 2b to identify the zeroth LL, which coincides with the Dirac point of the spectra taken at $B = 0\ T$ and exhibits a distinctive broadening.[19, 26] The remaining negatively dispersing peaks are indexed $LL_{-1}, LL_{-2}, \ldots, LL_N$ with respect to $LL_0$. For example, $LL_{-1}$ can be seen in the blue trace of Fig. 2c.

In Fig. 4b we plot the relative energy of the LL peaks $|E_{LLN} - E_{LL0}|$ as a function of $\sqrt{|N|B}$, where $N$ is the peak index and $B = 8$ T (see Supporting Information for more details on peak extraction). The data's linearity confirms the expected behavior of graphene's LLs and bolsters our analysis procedure. Figure 4b also includes relative peak energies taken at two additional values of $n$. While all three datasets are linear, their slopes noticeably increase as $|n|$ approaches zero, suggesting that the $v_F$ is $n$ dependent (see Supporting Information for $v_F$ vs $n$ at numerous $n$ values). Such behavior was reported previously in STS studies of graphene/hBN heterostructures and was attributed to electron-electron interaction effects.[24] Further studies are necessary to unambiguously determine if the observed $n$ dependence of $v_F$ in our experiment is also related to electron-electron interactions.

Finally, we use sheared $dI/dV_b(V_b, n)$ maps to examine how graphene's LLs fill and empty with charge at high magnetic fields. Tunneling spectra taken at $B = 18\,T$ presented in Fig. 3c show a well-defined staircase pattern. Similar patterns have been reported previously in two-dimensional electron systems,[7] and in STS studies of graphene[8,9], though not at $B = 18T$. These staircase patterns arise from highly compressible LLs for which the Fermi level is pinned at a constant energy as they are filled with charge. Once a LL is completely filled the introduction of additional charge causes the Fermi level to rapidly jump to the next LL. The resulting tunneling map shows a LL staircase, where the length of each stair ($\Delta n$) is equal to the number of degenerate cyclotron orbits in a given LL. Here we extract $\Delta n$ for different $B$ and plot the corresponding points in Fig. 4c. Without any fitting parameters, the data show excellent agreement with the theoretical degeneracy $\Delta n = gB/\phi_0$, where $\phi_0 = \frac{h}{e}$ is the flux quantum, and $g = 4$ is the expected single-particle LL degeneracy due to valleys and spins.

In conclusion, we developed a systematic method for interpreting and analyzing the planar tunneling spectra acquired from graphene TFETs. Our study yields direct identification of graphene's electronic structure features near and far from the Fermi level as they fill and empty with charge. By shearing $dI/dV_b(V_b, V_G)$ to attain $dI/dV_b(V_b, n)$ we resolved the Dirac point and numerous LLs at varying $n$. For $B = 0\,\text{T}$ we found remarkable agreement between our measurement of the Dirac point and single particle considerations. At finite $B$ we used our measurements of LLs and tight binding considerations to extract $v_F$ for three different *n* values. Although each extracted $v_F$ was on the order of the expected value for graphene, clear differences were apparent among these $v_F$ values. Such differences may arise from electron-electron interactions. Finally, we performed planar tunneling spectroscopy of graphene at unprecedented magnetic fields and observed well defined LL staircases. From these staircases, we extract a LL degeneracy that shows excellent agreement with single particle theory. We anticipate that the TFET devices presented here will be compatible with yet higher magnetic fields such as 45T, the highest constant magnetic field currently available. Thus, our TFET devices and shearing analysis technique will enable direct and thorough characterization of graphene's (and other 2D materials') electronic structure as it is altered by electron-electron interactions.

**Acknowledgments:** We thank S. Tomarken for insightful discussions on tunneling spectroscopy, F. Joucken and Z. Schlesinger for helpful feedback. We also thank D. Lederman for providing the low temperature cryostat and cryostat operation expertise that facilitated data acquisition for this work.

**Author contributions:** J.V.J. and J.L.D. conceived the work and designed the research strategy. J.L.D. and J.V.J. performed data analysis. J.L.D., Z.G., J.Y., and E.A.Q. fabricated the samples under J.V.J.'s supervision. K.H. and S.C. facilitated sample fabrication. K.W. and T.T. provided the hBN crystals. J.L.D., S.M., Z.L., D.S., P.G.L., and A.P.R. facilitated low temperature measurements. J.L.D., Z.G., C.N.L., and J.V.J carried out tunneling spectroscopy measurements. J.L.D., Z.G., and J.V.J. formulated the theoretical model. J.V.J. and J.L.D. co-wrote the paper. All authors discussed the paper and commented on the manuscript.

**Funding Sources:** J.V.J acknowledges support from UCOP and from the Army Research Office under contract W911NF-17-1-0473. A.P.R. acknowledges support from U.S. Department of Energy grant DE-SC001786. K.W. and T.T. acknowledge support from the Elemental Strategy Initiative conducted by the MEXT, Japan and the CREST (JPMJCR15F3), JST. A portion of this work was performed at the National High Magnetic Field Laboratory, which is supported by the National Science Foundation Cooperative Agreement No. DMR-1644779* and the State of Florida.

**Figure 1 | Tunneling field effect transistor (TFET) schematic and gate tunable tunneling spectroscopy dI/dV$_b$ at zero magnetic field. (a)** Graphene (gray) sandwiched between two hexagonal boron nitride (hBN) flakes, both blue. This stack is supported on a SiO$_2$/Si wafer (lavender/purple), which enables application of a gate voltage ($V_G$). A gold electrode (yellow) rests on top of the hBN/graphene/hBN stack and enables application of a voltage ($V_{probe}$) for inducing a tunneling current. **(b)** Optical micrograph of (a) with graphene (outlined in black) sandwiched between a thin (1.2 nm) hBN tunneling barrier (outlined in red) and a thick (20 nm) hBN supporting substrate. Six Cr / Au electrodes lie atop the thin hBN and V$_p$ is applied to the probes individually to perform tunneling measurements. The seventh electrode on the far-right contacts graphene directly. **(c)** $dI/dV_b(V_b, V_G)$ tunneling conductance map of graphene at B = 0 T and T = 4.2 K. **(d)** Numerical derivative of (a) taken with respect to $V_G$. Spectral features (i) and (ii) discussed in the main text are marked with black dotted and dashed lines, respectively. **(e)** $dI/dV_b(V_b)$ spectra taken along the colored diagonal lines in (a). Each line corresponds to different graphene charge densities *n,* listed in the legend. Line profiles are vertically staggered for clarity. Black arrows indicate the energy of feature (ii), which shifts with different *n*.

**Figure 2 | Gate tunable tunneling spectroscopy dI/dV$_b$ with well-defined *n*. (a)** Sheared $dI/dV_b(V_b, n)$ map of a graphene TFET, where charge density $n = C_{tg}V_b + C_{bg}V_G$ and $C_{tg}$, $C_{bg}$ are the graphene-tunneling probe and graphene-back gate capacitances, respectively. The shear accounts for the top gating effect of the tunneling probe, as discussed in the main text. **(b)** Sheared $dI/dV_b(V_b, n)$ map of the same graphene TFET for B = 8 T. The horizontal red line indicates the Fermi level of graphene $\mu_g$, white brackets coincide with filled and empty electron states below and above $\mu_g$, respectively. **(c)** Comparison of the $dI/dV_b(V_b)$ spectra taken at charge density $n = -1.8 \times 10^{12}$ cm$^{-2}$ when $B = 0$ T (blue) and $B = 8$ T (green). These spectra correspond to the

dashed vertical line cuts in (a) and (b) with respective colors. The spectra are vertically offset for clarity.

**Figure 3 | Gate tunable tunneling spectroscopy dI/dV$_b$ with well-defined *n* at high magnetic fields. (a-c)** Sheared $dI/dV_b(V_b, n)$ map of a graphene TFET for B = 8 T (a), B = 13 T (b) and B = 18 T (c). The former was performed at T = 4.2 K, while the latter two were performed at T = 0.5 K. Landau level degeneracy $\Delta n$, indicated in (c) and discussed in the main text. **(d)** Comparison of the $dI/dV_b(V_b)$ spectra taken along the charge neutral line of $dI/dV_b(V_b, n)$ maps from (a-c), with line colors corresponding to the color of the dashed vertical line in each of the maps. Spectra are vertically offset for clarity with the orange trace taken for B = 8 T, blue trace taken for B = 13 T, and black trace taken for B = 18 T. Pink arrows indicate the prominent LL$_{-1}$, which shifts away from $V_b = 0$ as B increases.

**Figure 4 | Energy dispersion of the Dirac point, Landau level energies, and state degeneracy factor. (a)** Energy of the Dirac Point in graphene with varying charge density. The position of spectral feature (ii) in $V_b$ (which is denoted as a dashed line in Fig. 2b) is plotted against *n*. The phonon energy $\hbar\omega = 67$ meV has been subtracted from the measured position of spectral feature (ii). Such data from two separate devices (denoted as red circles and blue diamonds) are compared with the theory $E_D = \hbar v_F \sqrt{n\pi}$, here $v_F = 10^6$ m/s. **(b)** Determination of the Fermi velocity at different *n*. Landau level (LL) peak energies are plotted against the square root of the LL index *N* and *B*, where B = 8 T. Error bars in LL position are smaller than marker sizes. Linear fits are used to extract the Fermi velocity at each of the three different *n*, shown in the legend. **(c)** Degeneracy of LL$_{-1}$ at different magnetic fields. The black line shows the expected degeneracy $\Delta n = g\left(\frac{B}{\phi_0}\right)$ where $g = 4$ is the single-particle degeneracy factor for graphene and $\phi_0 = \frac{h}{e}$ is the flux quantum.

Figure 1

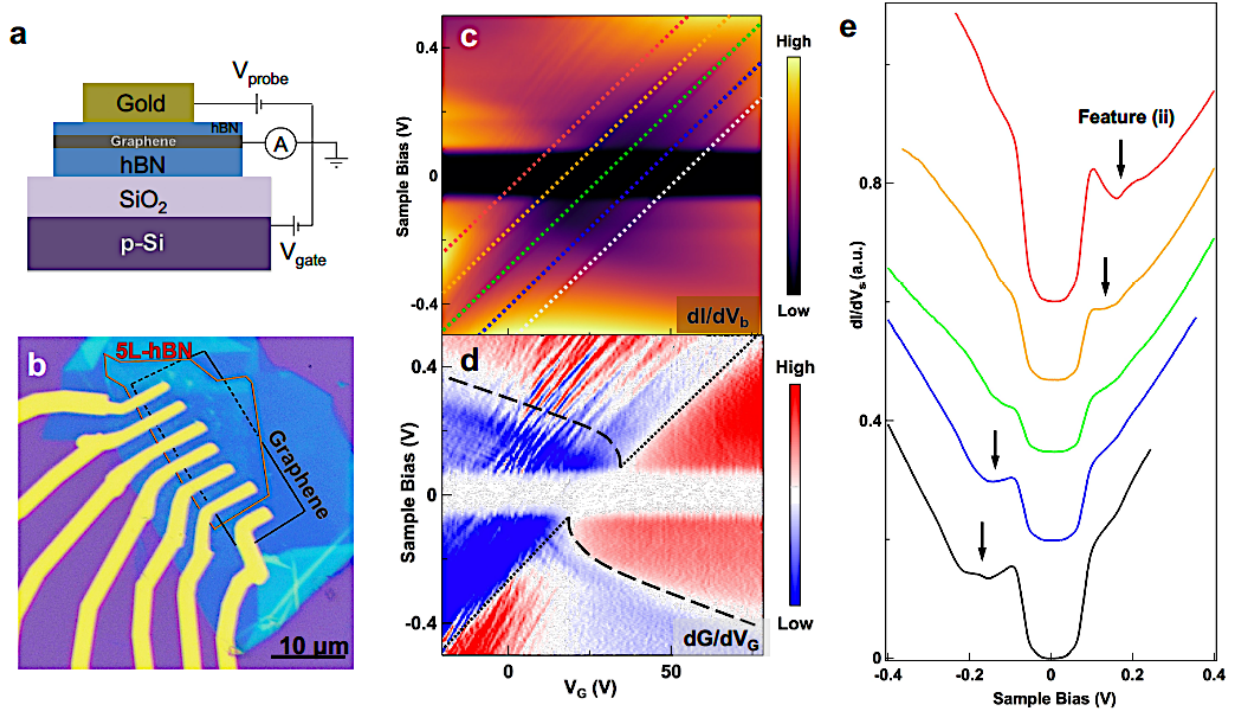

# Figure 2

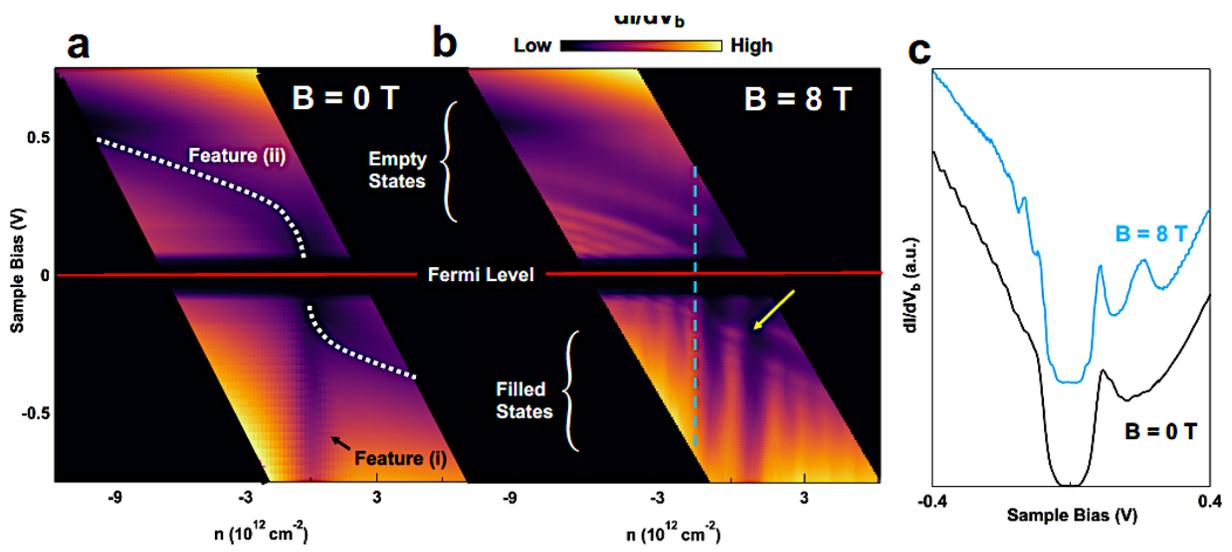

Figure 3

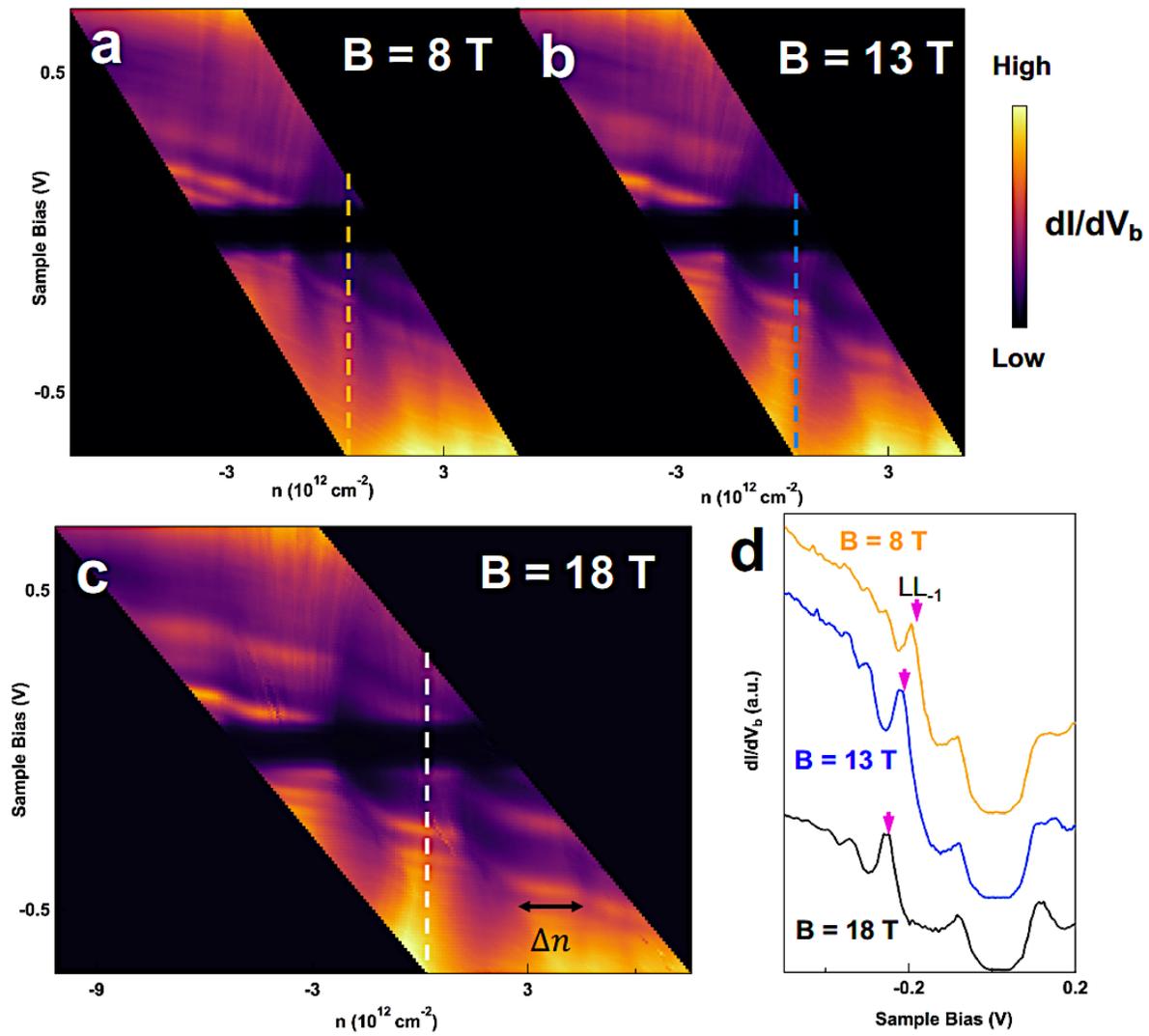

# Figure 4

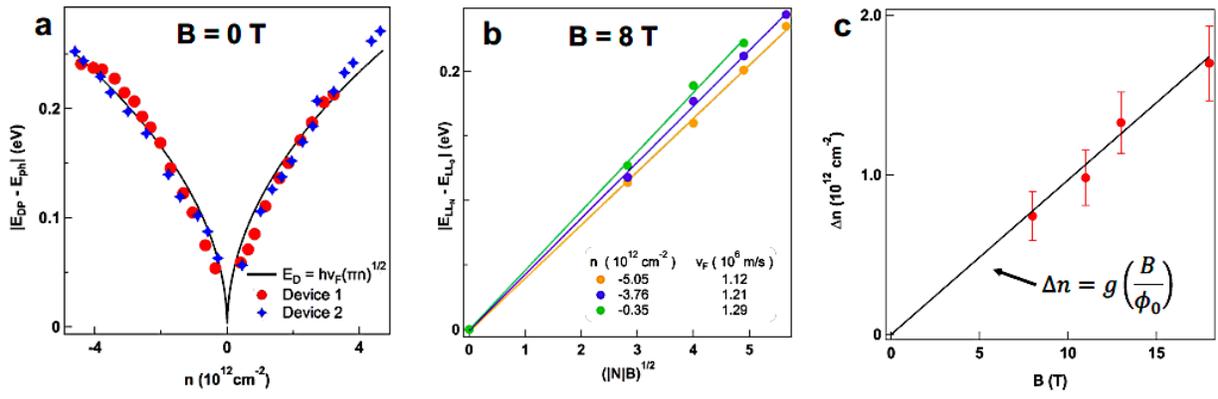